\documentclass{article}
\usepackage{ijcai25}

\usepackage{times}
\usepackage{soul}
\usepackage{url}
\usepackage[hidelinks]{hyperref}
\usepackage[utf8]{inputenc}
\usepackage[small]{caption}
\usepackage{graphicx}
\usepackage{amsmath}
\usepackage{amsthm}
\usepackage{booktabs}
\usepackage{algorithm}
\usepackage{algorithmic}
\usepackage[switch]{lineno}
\usepackage[capitalize]{cleveref}
\usepackage{subfig}
\usepackage{stfloats}
\usepackage{balance} % for balancing columns on the final page
\usepackage{microtype}
\usepackage{placeins}
\title{Implicitly Aligning Humans and Autonomous Agents\\ through Shared Task Abstractions}

\author{
St\'ephane Aroca-Ouellette$^1$
\and
Miguel Aroca-Ouellette\and\\
Katharina von der Wense$^{1,2}$\And
Alessandro Roncone$^1$\\
\affiliations
$^1$University of Colorado Boulder\\
$^2$Johannes-Gutenberg Universität Mainz\\
\emails
\{first.second\}@colorado.edu (omit accents, include hyphens),
}

\begin{document}

\maketitle

\begin{abstract}
In collaborative tasks, autonomous agents fall short of humans in their capability to quickly adapt to new and unfamiliar teammates. We posit that a limiting factor for zero-shot coordination is the lack of shared task abstractions, a mechanism humans rely on to implicitly align with teammates. To address this gap, we introduce HA$^2$: Hierarchical Ad Hoc Agents, a framework leveraging hierarchical reinforcement learning to mimic the structured approach humans use in collaboration. We evaluate HA$^2$ in the Overcooked environment, demonstrating statistically significant improvement over existing baselines when paired with both unseen agents and humans, providing better resilience to environmental shifts, and outperforming all state-of-the-art methods. Code can be found \href{https://github.com/HIRO-group/HA2/}{here}\footnote{\url{https://github.com/HIRO-group/HA2/}}
\end{abstract}

\section{Introduction}

Successful collaboration requires individuals to efficiently adapt to new teammates. This capability, often referred to as ad hoc teaming \cite{PLASTIC} or zero-shot coordination \cite{otherplay}, is an area where humans consistently outperform state-of-the-art autonomous agents. We argue that this disparity arises because humans have access to shared task abstractions \cite{hta_ext}, which provide a common foundation that facilitates seamless, implicit coordination. In this paper, we argue that maximizing an agent's ability to collaborate with humans requires providing them with shared task structures and demonstrate the effectiveness of this approach through a large-scale human study.

% With the rising prevalence of autonomous agents, an increased emphasis has been placed on how these agents interact with humans. 
%As autonomous agents become increasingly pervasive, it is important to understand and optimize how these agents interact with humans \textit{on human terms}.
%Given the vast landscape of human cognition, abilities, and preferences, successful Human-AI Teaming requires agents to rapidly adapt to new and previously unseen teammates, as well as evolving environmental conditions.
%This adaptability is 
% Unfortunately, conventional methods such as self-play \cite{sp}, originally designed for competitive settings,
%Creating these collaborative agents has proven challenging for traditional reinforcement learning methods such as self-play \cite{sp}. These methods, designed for competitive scenarios, 
%prove inadequate for cooperative tasks as their generated agents are too rigid for human-agent teamwork.
%are ill-equipped to handle the challenges of human-agent teamwork.

\begin{figure}[t]
\centering
\includegraphics[width=\columnwidth]{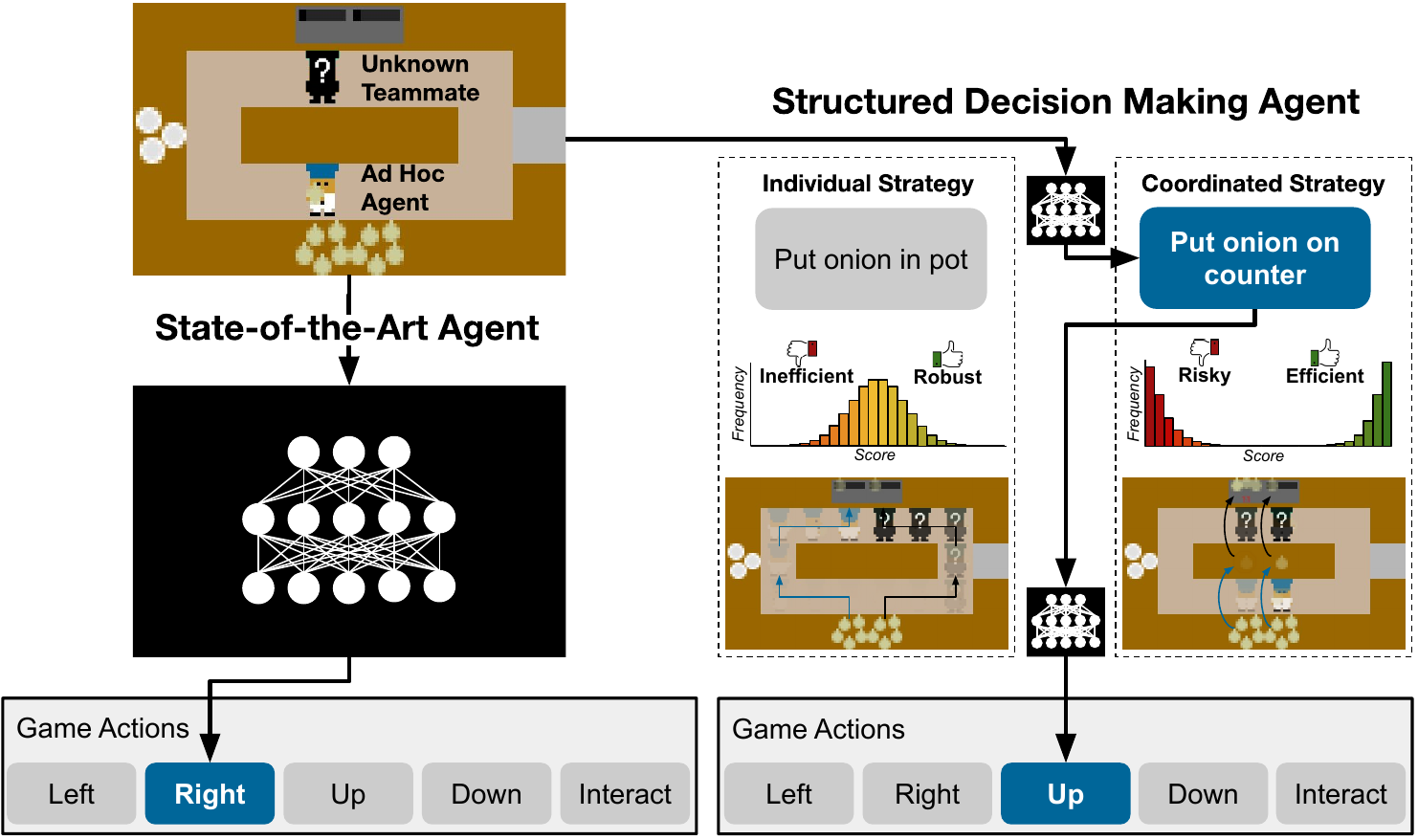}\small%
\caption{Depicted is a scenario in the Overcooked game where an agent is working with a new teammate. The agent could choose to play a coordinated strategy that is more efficient than an alternative individual strategy, but runs the risk of failure if cooperation is not achieved. Successful ad hoc teaming requires not only being able to perform multiple strategies, but also know when to apply the different strategies. Current SotA approaches subsume these decisions into a single black-box; in contrast, we propose that a structured approach to these decisions provides significant benefits.}%
\label{fig:example}%
\end{figure}

To elucidate the intricacies and challenges of zero-shot coordination between humans and agents, let's analyze a scenario in the collaborative game ``Overcooked'', in which players serve as many soups as possible within a time limit. In \cref{fig:example}, the ad hoc agent is playing with an unfamiliar teammate and must decide how to act next. One option, which we define as the `\textsl{individual}' strategy, involves obtaining an onion and placing it directly into the pot. This behavior is conservative but suboptimal: it can achieve a moderate score with any player, but will never achieve a top score.
Conversely, the agent could opt for a `\textsl{coordinated}' strategy, where the blue agent passes onions on the middle counter, hoping their teammate moves them from the counter to a pot. This strategy is more efficient since it eliminates the long walk around the kitchen, but it carries risk as success hinges on both agents adhering to the strategy. This example illustrates a key challenge in zero-shot teaming:
an agent must not only \textsl{acquire multiple distinct behaviors}, it also needs to \textsl{be able to quickly identify which behavior is most suitable for its teammate's skill level}.

Recent works have tried to overcome the challenges of rapidly adapting to new teammates by leveraging teammates of diverse capabilities. %varied training partners,
Agents have been trained with teammates that emulate human behavior \cite{oai} or with a population of teammates that varying levels of proficiency \cite{fcp,anyplay,lou2023pecan,gamma}. 
However, these approaches sidestep a key factor for achieving genuinely collaborative interaction; as depicted in \cref{fig:example}, current state-of-the-art systems consolidate high-level strategy decisions and low-level movement decision into one black box model. In contrast, humans are known to leverage hierarchical frameworks for cognitive processing \cite{hier_human_decision_making}, task management \cite{hta,managingtasks}, and human-human coordination \cite{hta_ext}, and human-robot collaboration \cite{roncone2017transparent,mangin2022helpful}. Further, it has been proposed that structured hierarchies are a core component of the human ability for fast generalization \cite{mind_hier}. In this paper, we design autonomous agents equipped with hierarchical structures that provide shared task abstractions that enable more efficient alignment with humans. 
These structures enable agents to focus on the most relevant information for the respective level of abstraction, prevent agents from overfitting to specific training patterns, and create task-oriented agents who may be more understandable to humans. While the benefits of shared task hierarchies are well-established in certain research domains \cite{saycan,voyager,hta}, state-of-the-art methods in human-agent interaction \cite{lou2023pecan,gamma} have yet to capitalize on this critical concept. This paper addresses this significant gap and advocates that shared task hierarchies should play a central role in human-agent interaction. Our findings show that leveraging shared task hierarchies can provide greater improvements compared to increasing diversity of training agents (cf. \cref{sec:other_work}).

In all, we present \textsl{Hierarchical Ad Hoc Agents} (\textsl{HA$^2$}), a method that leverages hierarchical reinforcement learning (HRL) to equip an agent with both low-level, efficient maneuvering behaviors and high-level, team-oriented strategies for effective synchronization with human teammates (cf. \cref{fig:arch}).
Importantly, HA$^2$ is agnostic to the underlying training algorithms, and serves as an augmentative layer that complements state-of-the-art methods (SotA, e.g., \cite{fcp,oai,lou2023pecan,gamma}), leading to statistically significant improvements. Further, this is a deeply generalizable method, as humans have demonstrated the ability to create tasks hierarchies across a broad range of human-human collaborative tasks \cite{hta_ext}. Through extensive evaluations, we find that HA$^2$%\footnote{Code will be open sourced at the end of the anonymity period.} 
offers statistically significant advantages with highlighted by the following contributions:
1) HA$^2$ outperforms all baselines by over $18.0$\% when paired with a set of unseen agents, and  
2) by over $18.3$\% when paired with humans. Moreover, 3) HA$^2$ is significantly preferred by humans, and found to be more fluent, trusted, and cooperative than baselines.
To further test the generalizability provided by hierarchical structures, we test the agents zero-shot on modified versions of the game layouts and show that 
4) HA$^2$ is more robust environmental changes, outperforming baselines by more than 10.5x on these layouts.

\section{Related Works}
\textbf{\textit{Zero-Shot Coordination}}
% As the ability of autonomous agents has grown, there has been 
The fast-evolving landscape of deep RL agents that interact in the real world has prompted
increased investigation into how they can and should interact with humans \cite{survey,survey2}. A critical challenge is the development of agents capable of zero-shot coordination with human partners \cite{ad-hoc-motivation}.

\textbf{\textit{Training partners}}
Prior research identified the limitations of agents trained via self-play---most notably, their behavioral rigidity. To address this, work has enhanced self-play through robust strategy discovery \cite{otherplay,klevel,sarkar2023diverse}, off-belief learning \cite{offbelief}, or training with a population of pretrained agents \cite{env-partner-diversity-overcooked}. 
% Within the research of pretrained populations, research has shown the benefits of using teammates that resemble human behavior \cite{oai} (modelled through imitation learning)
Notable advances were made with the use of teammates trained via imitation learning \cite{oai}%
, that vary in ability \cite{fcp}, or are specifically trained to be diverse \cite{anyplay,mep,trajedi}. 
More recent work has investigated ensembling training partners to create a richer diversity without additional computational costs \cite{lou2023pecan}. 
Our work is directly compatible with this body of work by augmenting the agents with human-aligned structures.

\noindent\textbf{\textit{Intention Prediction and Planners}}
A different line of research has focused on modeling the teammate's intention prediction for collaborative tasks \cite{ad-hoc-task-learning,intent-recognition}. 
This work often leverages online planners, which have nice properties for ad hoc agents within certain restrictions \cite{online-planning-ad-hoc}. 
Although \cite{bayesian,hier-overcooked} employ hierarchical structures in their planners, they lack real-world applicability due to computational constraints for complex environments and have not been tested with real human game-play. 
% . However, both use planners as their underlying method for action selection, which has several drawbacks. Primarily, it limits the environments they can be used on, as the planning time on even slightly complex environments can be much too slow for real-time play. 
\cite{oai} compared their models to planning-based methods, but were only able to use planners in two of their fivelayouts. When playing any agent for which they did not have an accurate model of (e.g., a human), performance dropped dramatically. 
% Crucially, neither \cite{bayesian-overcooked} or \cite{hier-overcooked} ran any trials with real humans.
Though out-of-scope here, we believe that modeling would compliment our proposed method.

\noindent\textbf{\textit{Type-based Agents}}
Ad hoc teaming has also been investigated by using type-based agents that rely on a pre-generated population of diverse teammates. 
The PLASTIC framework \cite{PLASTIC} offers two strategies: 
% PLASTIC-Model models previous teammates and uses the most similar model to the human to plan the next action. PLASTIC-Policy 
PLASTIC-Model, which employs the most human-like teammate for action planning, and PLASTIC-Policy, which 
first learns then selects the most appropriate complementary policy for each teammate.
% A similar approach to PLASTIC-Policy is used in \cite{hats}, however they use a different metric to determine similarity between humans and the existing population.
This latter approach is paralleled in \cite{hats}, albeit with a distinct similarity metric.
Finally, \cite{AATeam} %subsumes the teammates into several world models that they learn policies for. 
takes this further by subsuming teammates into world models, and using them to learn respective policies.
%During inference, %they choose which world model and associated policy to use based on the human's actions.
%the agent selects the most appropriate world model.% based on observed behavior.

%\subsubsection{Other Approaches}
% There is a line of research investigating game-theoretic approaches for ad hoc teaming \cite{game-theory-book,game_formulation,best-response-learn}
%Game-theoretic models have also been explored for ad hoc teaming \cite{game-theory-book,game_formulation,best-response-learn}%
%, as well as work investigating the carry over of training strategies that are effective for human-human teams \cite{crosstraining}.

\noindent\textbf{\textit{Hierarchical Reinforcement Learning}}
As breaking down complex tasks into sub-tasks is used in many facets of life, HRL is a well-studied area \cite{options,MAXQ,feudal-rl,fun}.
HRL has also been extended to multi-agent cooperation, either by deploying a central manager to oversee multiple agents \cite{feudal-rl-COOP}, or by imbuing each agent with its own hierarchical architecture \cite{multi-agent-hrl,multi-agent-hrl-v2}.
Other work has ventured into learning the sub-tasks \cite{skill-discovery,role-learning}.
Most similar to our work is HiPT \cite{hipt}. 
However, the work differs in several critical ways: 
1) Our motivation stems from aligning structures between humans and agents; thus, our abstracted layer between Worker and Manager is fully human interpretable.
2) Our method consistently outperform HiPT across all layouts. See \cref{sec:other_work} for comparative results.
3) We show that our architecture enables greater generalizeability to shifts in the game layouts, a feature not shown in HiPT.
4) We show that HA$^2$ provides significant benefits regardless of which training teammates are used.

% \label{sec:layouts}
% \begin{figure*}
% \centering
% \subfloat[\centering Asymmetric Advantages (AA)]{{\includegraphics[height=2.9cm]{figs/asymmetric_advantages.pdf} }}
% \subfloat[\centering Coordination Ring (CoR)]{{\includegraphics[height=2.9cm]{figs/coordination_ring.pdf} }}
% \subfloat[\centering Counter Circuit (CC)]{{\includegraphics[height=2.9cm]{figs/counter_circuit.pdf} }}
% \subfloat[\centering Cramped Room (CrR)]{{\includegraphics[height=2.9cm]{figs/cramped_room.pdf} }}
% \subfloat[\centering Forced Coordination (FC)]{{\includegraphics[height=2.9cm]{figs/forced_coordination.pdf} }}
% \caption{The five Overcooked layouts used. From \protect\cite{oai}.}
% \label{fig:5envs}
% \end{figure*}

\section{Method}

%the baselines we compare against (\cref{baselines}), and how we extend the baselines using our hierarchical agent (\cref{hrl}).{\color{red}remember to fix when you remove the baselines}

\subsection{Environment} \label{sec:env}
Following prior work in zero-shot human-AI teaming \cite{oai,fcp,haha}, we study the use of hierarchical structures using all five layouts in the Overcooked environment developed by \cite{oai}.
The goal of this collaborative game is to serve as many soups as possible in the time limit. To accomplish this, players must perform a sequence of task from retrieving onions and placing them in a pot to serving completed soups. Upon service, the team is rewarded with $20$ points. 

At each timestep, each player can choose to move \{\textsc{\textit{up, down, left, right}}\}, \textsc{\textit{interact}} with an object (for picking up/placing/serving objects), or \textsc{\textit{stay}} still. To effectively play Overcooked, agents must both coordinate on high-level sub-tasks and low-level movement patterns. At the sub-task level, players should avoid redundant and inefficient sub-tasks such as each retrieving a dish if only one soup is cooking. At a low-level, players must be cautious to avoid collisions. 
This layered complexity makes Overcooked a particularly apt testbed for human-agent collaboration.

\subsection{Sub-tasks}\label{sec:env:subt}
% The first step in creating a hierarchical structure is to define meaningful sub-tasks. 
% Fortunately, 
% Overcooked naturally lends itself to a subdivision of the main objective---serving soups---into sub-tasks. 
In human-human game-play, synchronization typically occurs at this sub-task level.
In the Overcooked environment, sub-task identification is facilitated by the \textsc{\textit{interact}} action, which serves as a delineating event. Utilizing it, we enumerate all possible outcomes resulting from the 'interact' action to define our set of sub-tasks: (1) Pick up onion from onion dispenser (2) Pick up onion from counter (3) Pick up dish from dish dispenser (4) Pick up dish from counter (5) Place onion in pot (6) Place onion on counter (7) Get soup from pot (8) Place dish on counter (9) Get soup from pot (10) Place soup on counter (11) Serve soup (12) Unknown
% We note that it is straightforward to extend these sub-tasks to new ingredients by adding the following four sub-tasks per new ingredient: pick up <ingredient> from dispenser, pick up <ingredient> from counter, place <ingredient> in pot, place <ingredient> on counter.

\subsection{HA$^2$: Hierarchical Ad Hoc Agent} \label{hrl}
\begin{figure}[t]\centering
\includegraphics[width=\columnwidth, trim=4cm 4cm 4cm 4cm, clip]{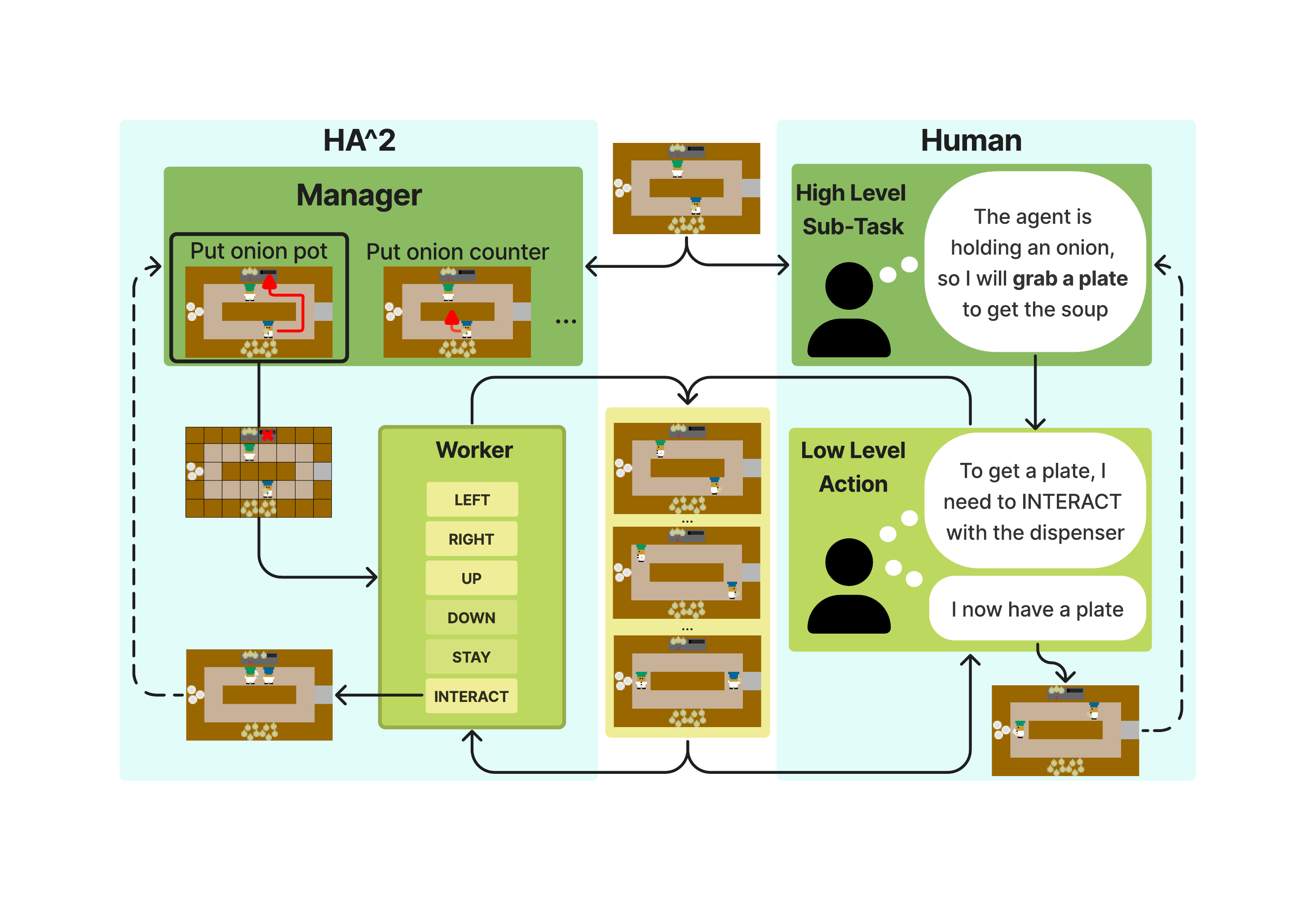}
\caption{An overview of the HA$^2$ architecture. Similar to human behavior, an observation is initially processed by the Manager to decide on the next high-level sub-task. Subsequently, the Worker executes the necessary low-level actions to complete the sub-task. }\label{fig:arch}
\end{figure}

% All the above methods produce agents that can only output low-level actions. 
 % With the motivation that aligning agents and humans at a cognitive and behavioral level will benefit human-agent collaboration, 
Inspired by the notion that cognitive and behavioral alignment between humans and agents enhances human-AI Teaming, we adapt FuN \cite{fun} to introduce HA$^2$: Hierarchical Ad Hoc Agents. HA$^2$ aims to approach human-agent teaming as one would approach human-human teaming: by developing a shared and mutually understandable task hierarchy.
%Our work takes inspiration from the insights that in an HRL framework goal setting can be decoupled from goal achievement with a focus of creating an intermediate representation that mirrors how a human might approach the game.
HA$^2$ (cf. \cref{fig:arch}) consists of two tiers of models: 
% a Worker, who learns the low-level actions required to efficiently navigate the environment and complete specific sub-tasks; and a Manager who learns which high-level subtask to perform in order to optimize team performance. 
a Worker, that focuses on the efficient execution of sub-tasks while avoiding collisions; and a Manager that focuses on high-level-task synchronization with its teammate.
This decoupled architecture not only facilitates collaboration by allowing the models to focus on the information at their level of abstraction, but it additionally streamlines both their learning processes.  
    % In this section, we first outline how we define a sub-task in Overcooked, then outline the Worker agent, and lastly outline the Manager agent.
    % \cref{fig:arch} shows a full overview of HA$^2$.
% We can then use these defined each sub-tasks to train a Worker. 

\textbf{The Worker} is tasked with learning how to complete sub-tasks.
With reference to \cref{sec:env:subt}, this consists of moving to a certain location with a specific orientation and \textsc{\textit{interact}}ing with the environment.
% We first identify that each sub-task consists of moving to a certain location and interacting with it.
With this in mind, we add a layer to the lossless observation developed by \cite{oai} that indicates the end locations of the current sub-task. For example, for the sub-task `put onion in pot', each non-full pot would be marked in this layer of the observation. We then create a modified versions of the original environment. In this environment, each episode is associated with a sub-task and runs until the agent performs the \textsc{\textit{interact}} action or times out. If the agent completes the correct sub-task, they receive a reward of +$1$, otherwise they receive a reward of -$1$. Certain sub-tasks offer additional small rewards for more optimal completion methods. For tasks involving placing objects on or picking up object from counters, an additional reward is added for the numbers of steps that can be saved by using that counter compared to moving from the agent's current location. For placing onions in pots there is an additional reward for placing onions in the pot that has more onions. When an episode ends, a new sub-task is sampled from the list of possible sub-tasks given the current state. The sub-tasks are sampled inversely proportionally to how often that sub-task has been used previously in training to get a more even coverage of sub-tasks during training. Once the horizon of the original environment is reached, the environment resets the state to the standard start state and a new sub-task is sampled. Due to its undefined nature, we omit the unknown sub-task at this stage.  

% \subsubsection{Manager}
% The final component of the hierarchical structure is the Manager. 
\textbf{The Manager} is responsible for deciding which sub-task should be completed next. Specifically, it is trained to output a distribution over sub-tasks. To train the Manager, we again create a modified version of the original environment. In this environment, the action space is one of the $12$ possible sub-tasks. If the manager selects the undefined sub-task, the additional observation layer passed to the worker is empty. Unlike the base environment, not all actions are possible for each state: for example, the agent cannot put an onion in the pot if they are not carrying an onion. To address this, we mask out all sub-tasks which are not possible at the current time-step. 
%, similar to \cite{maskablePPO}. 
Once the sub-task has been chosen, the associated observation is passed to the Worker, who selects the low-level action for that timestep. We found that having the Manager select sub-tasks at each time-step improved sample efficiency and overall performance given the computational budget. The reward structure is the same as the base environment with a reward of $20$ for each soup served.

\section{Experimental Design}
\subsection{Baselines and HA$^2$ models} \label{baselines}
We implement two baselines representative of the existing approaches in the field: Behavioral Cloning Play (BCP), \cite{oai} and Fictitious Co-Play (FCP) \cite{fcp}. BCP\footnote{BCP was originally named PPO$_{BC}$ in \cite{oai} and renamed by \cite{fcp} to BCP. We use BCP in this paper for succinctness.} was designed to have an RL agent learn how to play with the movement patterns of a human. To do this, a behavioral cloning (BC) model is first trained from human data, and then a RL agent is trained with the BC model as its teammate. 
FCP is designed to have agents learn to play with a wide range of teammates. It first learns a population of self-play agents who vary in architecture and seed. It then augments the population by using three versions of each agent: its random initialization, roughly midway through training, and after completing training. It then trains the FCP agent to play with the whole population of agents. 

We note that HA$^2$ serves as an architectural enhancement within the agent, and that the agent can be trained using any type of teammate.
To demonstrate the applicability of HA$^2$ to existing methods, we train two versions of HA$^2$. HA$^2$$_{BCP}$ is trained using a BC teammate and is directly comparable to BCP. HA$^2$$_{FCP}$ is trained using a FCP population and is directly comparable to FCP. We train five iterations of each of the four agents using different random seeds and report the mean and standard error across seeds. 

To train the BC models, we closely follow the implementation in \cite{oai}, using their feature encoding as observation as well as their provided data. 
We make two small changes which we found improves performance. First, we remove all time-steps where both agents perform the \textsc{\textit{stay}} action. Second, in the loss, we weigh each action inversely proportional to their frequency in the dataset.
Following \cite{oai}, We divide the data in half, and train two models. The better model is used as the human proxy, and the worse model as the BC model. We note that these two agents are the only agents where we train one model per layout. 

The RL agents train one model for all layouts and use the $7$x$7$ egocentric view developed by \cite{fcp}.
However, instead of the convolutional neural network (CNN) used in \cite{fcp}, we flatten the observation and pass it through a two-layer multilayer perceptron (MLP) as we found it outperforms a CNN. We experiment using recurrent networks, as in \cite{fcp}, but found they also underperform MLPs. We additionally experiment with frame stacking, which we found outperforms a Recurrent PPO, but underperforms the standard PPO approach.

The training population for the FCP agent and HA$^2$$_{FCP}$ consists of eight self-play agents that vary in seed, hidden dimension ($64$ and $256$), and whether or not they use frame stacking. When training the population, we found that agents learned on the different layouts at different rates. To maintain a good balance of skill levels for each layout, we use different middle checkpoints for each population agent for each layout, with the checkpoints corresponding to points closest to where the agent reaches half the highest score for that layout.

Each population agent was trained for $10$ million in-game steps and the BC agents were trained for $300$ epochs. HA$^2$ and the baselines train at different rates with HA$^2$ taking the longest to train since it requires two predictions --- one from the manager, the other from the worker --- at each timestep. To keep a fair comparison, we train each agent for $48$ hours using the same V100 GPU. For HA$^2$, we use $24$ hours for the Worker and $24$ hours for the Manager. The $48$ hours equate to $\sim$$119$ million timesteps for BCP, $\sim$$119$ million timesteps for FCP, and $\sim$66 million timesteps for HA$^2$ (31 million for the Worker and 35 million for the Manager). We note that all agents reached over 98\% of their top performance within the first half of this training.

\subsection{Research Questions and Experiments}

\textbf{\textit{RQ1: Does HA$^2$ improve performance with unseen agents?}}
We hypothesize that the addition of a hierarchical structure will help the agent's models focus more closely to the salient information at their respective level of abstraction. Further, we hypothesize that it learn more general game concepts by preventing it from over-fitting to any specific training patterns. 
Since the reward is fully shared and because the agent can impact its training teammate's actions by influencing the observations, it follows that the agent will also maximize its actions to promote its teammates' high-scoring behaviors. 
When using low-level actions, this can quickly lead to weird specificities that generalize poorly---e.g. waiting to put the onion in the pot until the teammate is in a specific spot and facing a specific way. Enforcing a hierarchical structure should mitigates this effect since the Worker is not rewarded by teammate behaviors and the Manager has no control over the movement of the Worker. 

To test this initial hypothesis, we compare the performance of HA$^2$$_{BCP}$ and HA$^2$$_{FCP}$ to their respective baselines when paired with three agents of varying capability: a self-play model (fully distinct from any in the FCP population), the human proxy model, and an agent that performs random actions. See \cref{sec:rq1_results} for the results of this experiment.

\textbf{\textit{RQ2: Does HA$^2$ create higher performing and more fluent human-agent teams?}}
\label{sef:rq3}
The primary motivation for this work is to develop agents that are effective at collaborating with humans. 
Human teammates present unique challenges to autonomous agents---prime among them the fact that humans have a significantly higher ability to adapt. 
In turn, this requires agents paired with humans to not only being able to adapt themselves, but also to make it easy for a teammate to adapt to them. Beyond the improved generalizability we test for, we hypothesize that HA$^2$'s structure will make them more task-focused and in turn more understandable to humans.

To this end, we conduct an IRB-approved online user study. We use a within-subjects design for the study where each participant plays with two agents on each layout. To test our above hypothesis, we evaluate both objective performance and subjective preferences between pairs of agents. Each participant was first provided with an instruction page, before completing a short tutorial that required them to complete all the steps to serve a soup to move on. Each participant then played an $80$ second round ($400$ steps at $5$FPS) with each agent on one of the layouts. Between each round, the participants had to answer eight questions adapted from \cite{hri_qs} asking them how much they agreed or disagreed with statement on a 7-point likert scale. After each pair, they were asked to rank which of the two agents they preferred. They then repeated this process for the other four layouts. The order of the layouts and agent they played with first within each layout was randomized. The chef that the agent and human controlled were consistent between the two comparative agents, but randomized between layouts and participants.

For this research question, we run two pairwise comparisons: HA$^2$$_{BCP}$ vs. BCP and HA$^2$$_{FCP}$ vs. FCP. We recruit $50$ participants for the BCP comparison and $25$ participants for FCP comparison. We filter out any participant that did not complete the full trial. We additionally filter out any pair of rounds (i.e., comparing two agents on one layout) where the human performed fewer than five subtasks in either round. This leaves us with $47$ and $24$ participants respectively. We recruit all participants from prolific.com. Participants were compensated using a base rate of \$3.00 plus a bonus incentive of \$0.04 for each dish served. The average participant compensation for these two studies was \$15.79/hour.
% CU Boulder's  (protocol 22-0495) re-add when de-anonymizing
This human survey was approved an Institutional Review Board, indicating that it presented minimal risk to participants. All participants provided informed consent for the study. Results for these human studies are in \cref{sec:rq3_results}.

\textbf{\textit{RQ3: Can HA$^2$ agents generalize better to changes in the layouts?}}
Since the hierarchical structure we are using is intrinsic to Overcooked at large, we posit that HA$^2$ should not only generalize better to different agents, but also generalize better to shifts in the layouts. For this experiment, we create a modified version of each layout by swapping two tiles in each layout. Since we do not have any trained unseen agents on these layouts, we evaluate the HA$^2$s and their respective baselines on these modified layouts by teaming each agent with themselves. \cref{sec:rq2_results} shows results for this experiment.

\begin{figure*}[htbp!]
\centering
% \includesvg[inkscapelatex=false, width = \linewidth]{figs/main_graph_w_sig_w_human_score_ha2.svg}
\includegraphics[width=\linewidth]{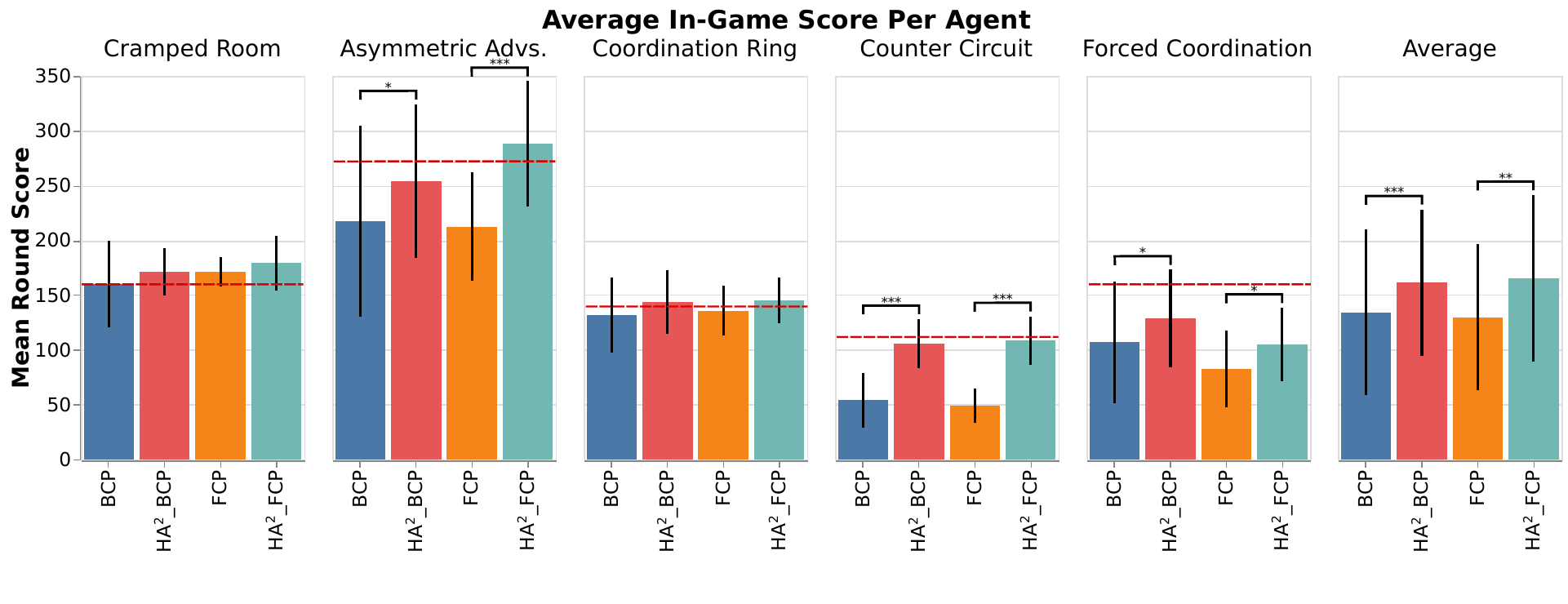}
\caption{Average score of HA$^2$s and the BCP and FCP baselines when paired with humans on each of the layouts. Each round was $80$ seconds long at $5$ FPS (T=$400$ steps). Significance markers: *=p$<0.05$, **=p$<0.005$, ***=p$<0.0005$. The red line indicates the max human-human score achieved on that layout from \protect\cite{oai} normalized to 400 steps.}
\label{fig:main_graph}
\end{figure*}

% We ran a third human study to compare a tuned version of HA$^2$ (HA$^2$$_{tuned}$) and an untuned version (HA$^2$). For this study, we recruited $35$ participants, which turned into $33$ participants after applying our filtering described above. We use the same compensation structure, and the average compensation for this study was \$15.98/hour. This study was also included in the IRB protocol. Results for the tuning study are shown in \cref{sec:rq3_results}.

\subsection{Significance Testing} 
\label{sec:sig}

For each pairwise comparison, we perform t-tests to measure significance. For the significance of team performance, we compare the score achieved directly.
% human--agent teams, HA$^2$$_{BCP}$ performed significantly better than BCP (p$<0.0162$), and HA$^2$$_{FCP}$ performed significantly better than FCP  (p$<0.0014$). 
For the ranking significance, we mapped every instance where an agent was preferred over its counterpart to a score of 1 and every other instance to a score of 0. We then used these scores to perform the t-tests. For the Likert questions, we mapped each agreement level to a score between -3 (strong disagree) and 3 (strong agree), with the neutral score being 0. We normalize all participants scores to have a mean of 0 and then use these score for perform the t-tests.

\begin{table}[t]
    \centering
    \begin{center}
    \setlength\tabcolsep{3pt}
    \begin{tabular}{|c|c|c|c|c|} \hline 
          &BCP&HA$^2$$_{BCP}$&FCP&HA$^2$$_{FCP}$\\ \hline \hline 
         AA&199.9$\pm$8.0&278.3$\pm$6.3&210.8$\pm$40.0&\textbf{293.5$\pm$7.2}\\ 
         CoR&79.2$\pm$4.2&133.3$\pm$3.2&138.6$\pm$2.5&\textbf{147.6$\pm$0.8}\\ 
         CC&17.1$\pm$11.4&91.2$\pm$5.0&74.3$\pm$19.3 &\textbf{99.9$\pm$2.8}\\  
         CrR&143.1$\pm$13.8&177.7$\pm$4.1&183.9 $\pm$4.7 & \textbf{185.5$\pm$2.3}\\  
         FC&73.1$\pm$5.6&\textbf{77.6$\pm$3.5}&56.7 $\pm$ 4.1&58.4$\pm$4.8\\ \hline
         Avg. &102.5$\pm$4.5& 151.6$\pm$2.4&133.0$\pm$8.8&\textbf{157.0$\pm$1.3}\\ \hline \hline
         $\sim$ AA & 23.6$\pm$41.5& 157.2$\pm$40.4&  7.6$\pm$14.2& \textbf{208.0$\pm$28.1}\\  
         $\sim$ CoR& 11.6$\pm$11.4& \textbf{152.8$\pm$7.0}&  22.8$\pm$6.4& 143.2$\pm$12.6\\  
         $\sim$ CC& 2.0$\pm$2.5& 70.0$\pm$15.8&  9.2$\pm$14.5& \textbf{110.0$\pm$35.5}\\
         $\sim$ CrR& 5.6$\pm$2.9& \textbf{162.4$\pm$15.2}& 0.8$\pm$1.6& 154.8$\pm$36.8\\ 
         $\sim$ FC& 10.4$\pm$8.9& 17.2$\pm$31.5&  3.2$\pm$3.0 & \textbf{20.8$\pm$31.7}\\ \hline
         $\sim$ Avg. & 10.6$\pm$9.5 & 111.9$\pm$13.4 & 8.7$\pm$2.4 & \textbf{127.3$\pm$7.1} \\ \hline
    \end{tabular}
\caption{Mean$\pm$SE score across $5$ random training seeds for HA$^2$s and their respective baselines. The score of each trained agent is the average across 10 trials of T=$400$ steps with each teammate. In the original layouts, the teammates are an unseen self-play agent, the human proxy, and a random agent. In the modified layouts (denoted with $\sim$ ), the teammate is a copy of the acting agent.}
\label{tab:unseen_agents}
\end{center}
% \vskip -0.1in
\end{table}

\section{Results}
\subsection{Zero-shot Coordination with Unseen Agents}
\label{sec:rq1_results}

We first compare HA$^2$ to the baselines---BCP and FCP---on their ability to generalize to new unseen agents. The results in \cref{tab:unseen_agents} clearly demonstrate the improvement provided by the hierarchical structure, with the HA$^2$s outperforming their respective baselines on every layout. Using HA$^2$ afforded an improvement of $47.9$\% when using BCP, and an improvement of $18.0$\% when using FCP. HA$^2$$_{BCP}$ performs best on forced coordination, and HA$^2$$_{FCP}$ performs best on all the other layouts and overall. We discuss a possible cause of this in \cref{sec:rq2_results}. We additionally note that HA$^2$ is more robust to the random seed than the baselines, with a lower standard error on each layout across the 5 random seeds.

\subsection{Zero-shot Coordination with Humans}
\label{sec:rq3_results}

\begin{figure}[h!]
\centering
% \includesvg[inkscapelatex=false, width=\columnwidth]{figs/main_likert_w_sig_ha2.svg}
\includegraphics[width=\columnwidth]{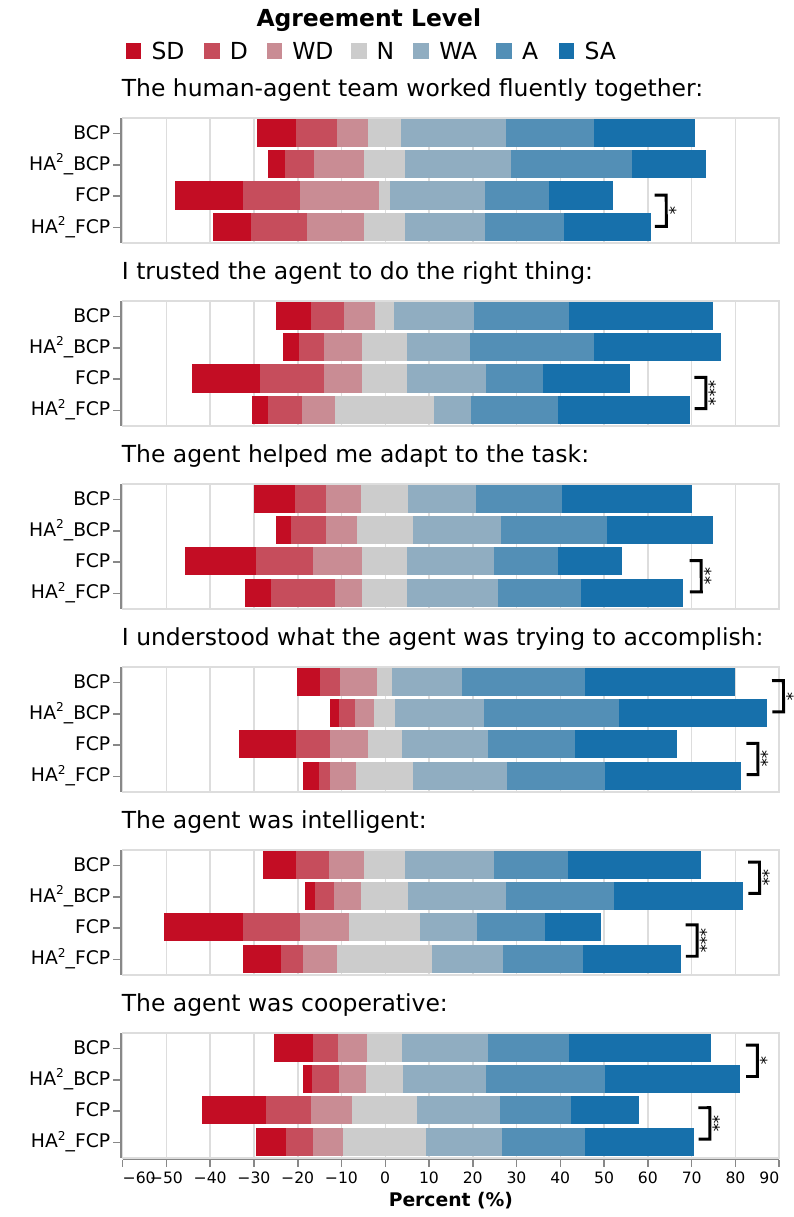}
\caption{Subset of results from the eight Likert-scale questions that participants answer after playing with each agent for the comparison between HA$^2$ and their baselines. Bars that are more blue indicate that people agree more strongly with the statement. Conversely, more red indicates that people disagreed more strongly with the statement. Significance markers: *=p$<0.05$, **=p$<0.005$, ***=p$<0.0005$. Legend: SD=Strongly Disagree, D=Disagree, WD=Weakly Disagree, N=Neutral, WA=Weakly Agree, A=Agree, SA=Strongly Agree.}\label{fig:likert_baselines}
\end{figure}

We now present the findings of our human study comparing HA$^2$ and the baselines. Results in \cref{fig:main_graph} demonstrate that in both HA$^2$$_{BCP}$ and HA$^2$$_{FCP}$ significantly outperform their respective counterparts on the overall score achieved, and on the asymmetric advantages, counter circuit, and forced coordination layouts. We note that the scores of the baselines in cramped room and coordination ring are closer to the maximum human-human \footnote{From \cite{oai}'s data normalized to 400 timesteps.} score achieved (dotted red line), leaving less room for improvement. 
As such, we anticipated there would be a smaller variability on these layouts. 
\cref{tab:ranking} further shows that HA$^2$ was significantly preferred over their counterparts. In RQ3, we had hypothesized that HA$^2$s would improve human-agent teaming because they are easier to understand, and therefore easier to adapt to. \cref{fig:likert_baselines} supports this hypothesis and shows that in both comparisons of HA$^2$ to the baselines, humans rated the HA$^2$s as significantly more understandable, intelligent, and cooperative. In the case of FCP and HA$^2$$_{FCP}$, humans also found that HA$^2$ was significantly more fluent, trusted, and more helpful at helping the humans adapt to the task. These results strongly support using shared task hierarchies for human-agent collaboration.

In line with the results with unseen agents, forced coordination is the one layout where BCP and HA$^2$$_{BCP}$ outperform their FCP counter parts. We hypothesize that this is due to it being the only layout where having an untrained teammate blocks the agent's ability to earn a reward. Since a third of FCP's training population are untrained agents, FCP and HA$^2$$_{FCP}$ effectively lose a third of their training. The results in the appendix of \cite{fcp} support this hypothesis showing that forced coordination is least benefitted by FCP. This can likely be remedied by excluding the untrained partners in layouts where coordination is required to achieve a non-zero score.

\begin{table}[b]
    \centering
    %For the BCP and FCP comparisons, the preference is across all layouts; for the tuned comparison it is only for the counter circuit layout.}
    \begin{tabular}{|l|c|c|} \hline 
         &  \% Preferred& p-value\\ \hline 
         HA$^2$$_{BCP}$ over BCP&  57.68& 0.0070\\ \hline 
         HA$^2$$_{FCP}$ over FCP&  65.25& 0.0000018\\ \hline 
         % HA$^2$$_{tuned}$ over HA$^2$&  66.67& 0.0093\\ \hline
    \end{tabular}
    \caption{Human preference between pairs of agents and their respective significance.}
    \label{tab:ranking}
\end{table}

Interestingly, when analyzing \cref{fig:main_graph,fig:likert_baselines}, we noticed that even if their overall scores were generally worse, BCP and HA$^2$$_{BCP}$ were better perceived on every subjective metric relative to their FCP and HA$^2$$_{FCP}$. This does pose the question of whether utilizing human behavior in training does provide a more human-like game-play, and in turn a more fluid experience for humans, which is supported by the results in \cite{gamma}.
We leave a more thorough investigation of this question, as well as the relationship between team performance and human perception, to future work.

\subsection{Comparison to State-of-the-Art}
\label{sec:other_work}

In \cref{tab:baselines}, we compare HA$^2$ to results published in other peer-reviewed work that uses the same overcooked environment. As each method employs a range of design decisions, this table should be viewed as a comparison of systems.
Notably, when paired with a human proxy, HA$^2$ is tied as the best performing agent, whereas when HA$^2$ is paired with real humans, HA$^2$ outperforms all other work by more than 23\%, showcasing HA$^2$'s adeptness at human collaboration. This is further emphasized when comparing to the most similar work of HiPT. HA$^2$ outperforms HiPT by 17\% when paired with a human proxy, and by 26\% when paired with real humans while using 15.1 times fewer timesteps (1 billion timesteps for HiPT vs $\sim$66 million timesteps for HA$^2$). 
This highlights HA$^2$'s greatest distinction from HiPT: \textsl{using \textbf{human-aligned} structures improves the training efficiency and performance of autonomous agents that collaborate with humans}. 
Lastly, we compare HA$^2$ to the most recent SotA method: GAMMA \cite{gamma}. Even with a simpler training population, HA$^2$ outperforms GAMMA by 25\% with a proxy human. Further, when paired with real humans on counter circuit, which is the only original layout on which they provide results with real humans, HA$^2$ outperforms the best version of GAMMA with a score of 110 compared to 91. In all, we show that shared task structures are a critical component when developing collaborative agents.

\subsection{Generalization to Shifts in Layouts}
\label{sec:rq2_results}

We report our results on the generalization ability of HA$^2$ and the baselines on the altered layouts. The latter half of \cref{tab:unseen_agents} shows that BCP and FCP overfit to the specific layouts and their performance drops dramatically when the layouts are changed. In contrast, the HA$^2$s are able to maintain a reasonable performance, and are over 10.5x better on the modified layouts. Together with the results in \cref{sec:rq1_results}, these results provide strong support for our hypothesis that the hierarchical structure enables the model to learn more generalizable concepts about collaboration and game-play.

% \subsection{Tuning HA$^2$}
% Lastly, we present the results comparing the tuned and untuned HA$^2$. While we found only a slight increase in average score when paired with humans --- from 104 to 108 --- humans significantly preferred the tuned agent (cf. \cref{tab:ranking}), and found it significantly more fluent, trustworthy, and intelligent (cf. \cref{fig:likert_tuning}). We believe that a reason for the lack of score increase is that the coordinated strategy is too complicated for inexperienced human players. To test this, we also paired HA$^2$ and HA$^2$$_{tuned}$ with themselves, and found tuning HA$^2$ improves the average score from $136.8_{\pm 6.3}$ to $154.0_{\pm 6.0}$, which is highest score achieved on this layout by any agent. For reference, the best SP agents achieve a score of 140.

\begin{table}[t]
    \centering
    \begin{tabular}{|l|c|c|c|} \hline 
         &  Training Steps&  W. Proxy& W. Humans\\ \hline 
         FCP &  1.0e9&  \textbf{157}& 119\\ \hline 
         MEP &  5.5e7*&  98& 98\\ \hline 
         TrajeDi &  5.5e7*&  76& 87\\ \hline 
         PECAN &  NR&  105& 134\\ \hline 
         HiPT & 1.0e9&  134& 131\\ \hline 
         GAMMA & 1.5e8 & 132 & NR\\ \hline 
         HA$^2$$_{FCP}$&  6.6e7&  \textbf{157}& \textbf{165}\\ \hline
    \end{tabular}
    \caption{Results comparing HA$^2$ to other published results. All results are taken from the respective works and adjusted to 400 timesteps, except for TrajeDi's results which are taken from \protect\cite{mep}. NR=not reported. * indicates that separate agents are trained for each layout and that the cumulative step count across layouts is presented. FCP \protect\cite{fcp}, MEP \protect\cite{mep}, TrajeDi \protect\cite{trajedi}, PECAN \protect\cite{lou2023pecan}, HiPT \protect\cite{hipt}, GAMMA \protect\cite{gamma},}
    \label{tab:baselines}
\end{table}

\section{Discussion}
\subsection{Summary}
% Throughout this paper, we have investigated the use of Hierarchical RL for effective human-agent collaboration. 
This paper offers a comprehensive investigation into the efficacy of leveraging shared task abstractions for enhancing human-agent collaborative systems. 
The experiments we conducted demonstrate that 
% when paired with either other unseen agents, or with real humans, our Hierarchical Ad Hoc Agent outperforms all baselines both on quantitative and qualitative metrics.
our Hierarchical Ad Hoc Agent (HA$^2$) significantly outpaces existing baselines in both quantitative and qualitative assessments.
In interactions with human participants, HA$^2$ created higher performing teams and was perceived by the human as more fluent, more understandable, more cooperative, and more intelligent. 
% We showed that HA$^2$ are not only able to generalize between different types of agents well, but can additionally generalize to shifts in layouts substantially better than the baselines.
Importantly, HA$^2$ displays robust generalization capabilities not only across diverse agent types but also in response to variations in environmental layouts---outperforming baseline models substantially in these regards.
Finally, we highlight how HA$^2$, even when trained using simpler training partners, outperforms all existing methods when paired with real humans.
%Lastly, we highlight HA$^2$'s ability to have its behavior adjusted post-training to produce new strategies that did not emerge during training.
% Importantly, we highlighted that HA$^2$'s design affords post-training behavior modification, and demonstrated how to induct novel strategies not acquired during training.
% Through HA$^2$, we demonstrate how hierarchical structures can create agents that are more robust to change and are more understandable to humans. 
These findings collectively highlight the utility of human-interpretable hierarchical structures in designing AI agents that are both resilient to changing conditions and intuitively collaborative \textsl{on human terms}. 
% We believe this foundation can provide a necessary step for safer and more efficient human-AI teams.
We posit that these advancements form a crucial building block toward more performant and efficient human-AI teams.

\subsection{Limitations}\label{sec:futurework}

%While we have demonstrated the value of a hierarchical structure for human-agent collaboration, there are drawbacks and limitations to the methods outlined here.
We now discuss the limitations of our proposed method.
% First, the hierarchical structure does come with an additional engineering burden. The action mask for the Manager has to be curated for each sub-task and certain layouts (e.g. in forced coordination, only one of the two agents can serve a soup, whereas in counter circuit both agents can).
The hierarchical structure in HA$^2$ necessitates additional engineering effort, both in the development of the structure, and the adjustments to the environment required to train the different modules. We note that the method in which to break-down large tasks into sub-tasks to create a hierarchy is not the focus of this work, and has been extensively explored in many domains including human factors research \cite{hta,hta_ext}, robotics \cite{saycan}, and single-agent long-horizon tasks \cite{voyager}. Rather, the focus of this work is demonstrating the importance of shared task hierarchies in human-agent collaboration.
% While existing reserach in HRL has shown the possibility of learning  sub-tasks  \cite{skill-discovery,role-learning, hipt}, a crucial component of our work is that the structure is interpretable by human. 
%Although research in HRL has shown the possibility of learning sub-tasks autonomously \cite{skill-discovery,role-learning,hipt}, our emphasis is on maintaining a framework that is intuitively understandable to humans. 
%However, humans already extensively apply this effort to break-down large tasks into sub-tasks for human-human teams \cite{hta,hta_ext}, indicating that the benefits are well worth the cost.
  
%Second, the work was conducted in an environment---Overcooked---that is characterized by well-delineated, discrete sub-tasks. 
% While in many situations the delineation between sub-tasks may be more difficult to define,
%While the demarcation between sub-tasks may require more nuance in other environments,
%human factors research has shown that virtually every task can be broken down into sub-tasks \cite{hta_ext}.
% and significantly more complicated problems have successfully been broken down into sub-tasks in the area of robotics \cite{saycan, code-as-policies}. 
%Moreover, more complex challenges in the field of robotics have been successfully disaggregated into sub-tasks \cite{saycan,code-as-policies}, lending credibility to the applicability and generalizability of our approach.

\subsection{Future Work}
We envision the following avenues for future work:  

First, to incorporate explicit mental models of teammate sub-tasks into agent planning, similar to \cite{ad-hoc-task-learning,intent-recognition}. We envision that these mental models will synergize with the abstracted manager sub-tasks allowing for more efficient computation of these models, and in turn providing the manager with an efficient understanding of human team members' capabilities and intentions.  

Second, we believe HA$^2$ shows promise as a framework to investigate human-agent communication in collaborative games; it is much easier to communicate at sub-task-level than at action-level. 

\section*{Ethical Statement}

%As with any technology, the application of the methods described in the paper can be used for good or bad.
We have shown that leveraging hierarchical structures can yield agents that collaborate more fluently and coherently with humans, potentially fostering greater trust and operational efficiency in human-AI teams.
% This can lead to increased trust and efficiency in human-agent teams.
% However, while HA$^2$ is more understandable than existing baselines, it still relies on neural networks, which are inherently opaque. 
However, it's crucial to recognize that, despite its improved transparency, the HA$^2$ paradigm is still underpinned by neural networks, which remain inherently opaque.
% People may be more likely to place undeserved trust into the system because they believe it is more trustworthy than it is. 
This opacity could induce misplaced trust in the system's capabilities or intentions.
% Lastly, as the agent is more controllable by biasing certain sub-tasks, malicious actors may use this power at specific times to endanger humans.
Additionally, the increased controllability of the agent through sub-task biasing introduces the potential for misuse, particularly by malicious actors aiming to compromise human safety.

\section*{Acknowledgments}
This work was supported by the Army Research Laboratory (Grants W911NF-21-2-02905, W911NF-21-2-0126) and by the Office of Naval Research (Grant N00014-22-1-2482).

\bibliographystyle{named}
\bibliography{ijcai25}
% \small

\end{document}